\documentclass[pra,aps,superscriptaddress,amssymb,amsmath,reprint,noeprint,floatfix]{revtex4-2}
\usepackage{natbib}
\usepackage{braket}
\usepackage{bm}
\usepackage{ulem}
\usepackage{siunitx}

\usepackage{graphicx}
\usepackage{hyperref}
\usepackage{xcolor}
\hypersetup{bookmarksnumbered}

\newcommand{\EqnOne}{
    \begin{equation}
        \begin{aligned}
        \chi=-\frac{\varepsilon_0 \omega}{2} \operatorname{Im}\left(\mathbf{E}^* \cdot \mathbf{B}\right),
        \end{aligned}
        \label{eqn:1}
    \end{equation}
}

\newcommand{\EqnTwo}{
    \begin{equation}
        \begin{gathered}
        g=\frac{2\left(A^\text{L}-A^\text{R}\right)}{A^\text{L}-A^\text{R}},
        \end{gathered}
        \label{eqn:2}
    \end{equation}
}

\newcommand{\EqnThree}{
    \begin{equation}
        \begin{gathered}
        A^{\text{L}, \text{R}}=\frac{\omega}{2} \operatorname{Im}\left(\mathbf{E}^* \cdot \mathbf{p}+\mathbf{B}^* \cdot \mathbf{m}\right),
        \end{gathered}
        \label{eqn:3}
    \end{equation}
}

\newcommand{\EqnFour}{
    \begin{equation}
        \begin{aligned}
            \mathbf{p}=\alpha_{\mathrm{e}} \mathbf{E}+i \kappa \mathbf{B}, \mathbf{m}=\alpha_{\mathrm{m}} \mathbf{B}-i \kappa \mathbf{E}.
        \end{aligned}
        \label{eqn:4}
    \end{equation}
}

\newcommand{\EqnFive}{
    \begin{equation}
        \begin{aligned}
           A^{\mathrm{L}, \mathrm{R}}=\frac{\omega}{2}\left(\alpha_e^{\prime \prime}|\mathbf{E}|^2+\alpha_m^{\prime \prime}|\mathbf{B}|^2\right) \mp \frac{2}{\varepsilon_0} \kappa^{\prime \prime} \chi,
        \end{aligned}
        \label{eqn:5}
    \end{equation}
}

\newcommand{\EqnSix}{
        \begin{equation}
        \begin{aligned}
        g=-\frac{2 \kappa^{\prime \prime}}{\omega \alpha_e^{\prime \prime}} \frac{\chi}{U_e},
        \end{aligned}
        \label{eqn:6}
    \end{equation}
}

\newcommand{\FigOne}{
    \begin{figure}[t!]
        \centering
        \includegraphics[width=0.9\linewidth]{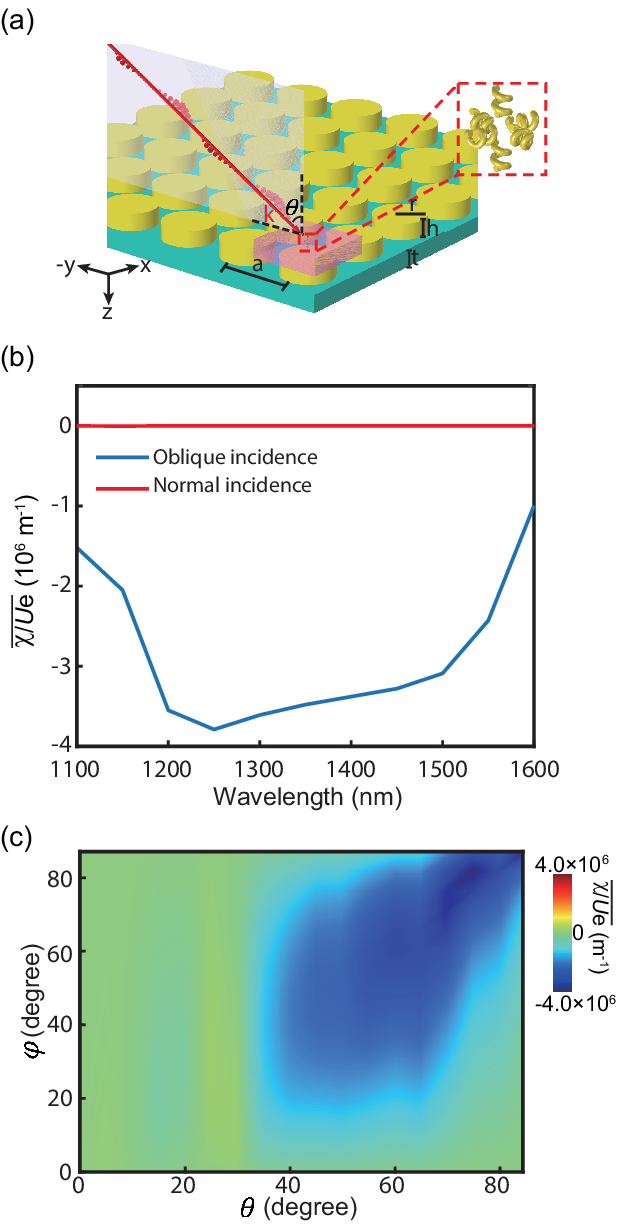}
        \caption{(a) Schematic of the achiral metasurface consisting of gold disks sitting on a silica substrate. The incident plane wave is linearly polarized with incident angle $\theta$ and polarization angle $\varphi$. The angle $\varphi$ is defined as the angle between the polarization direction and the \textit{yoz} plane. (b) Enhancement of the normalized optical chirality averaged over the red volume in (a). (c) Normalized optical chirality averaged over the red volume in (a) as a function of incident angle $\theta$ and polarization angle $\varphi$. The wavelength is set to be 1400 nm.}
        \label{fig:1}
    \end{figure}
}

\newcommand{\FigTwo}{
    \begin{figure}[t!]
        \centering
        \includegraphics[width=\linewidth]{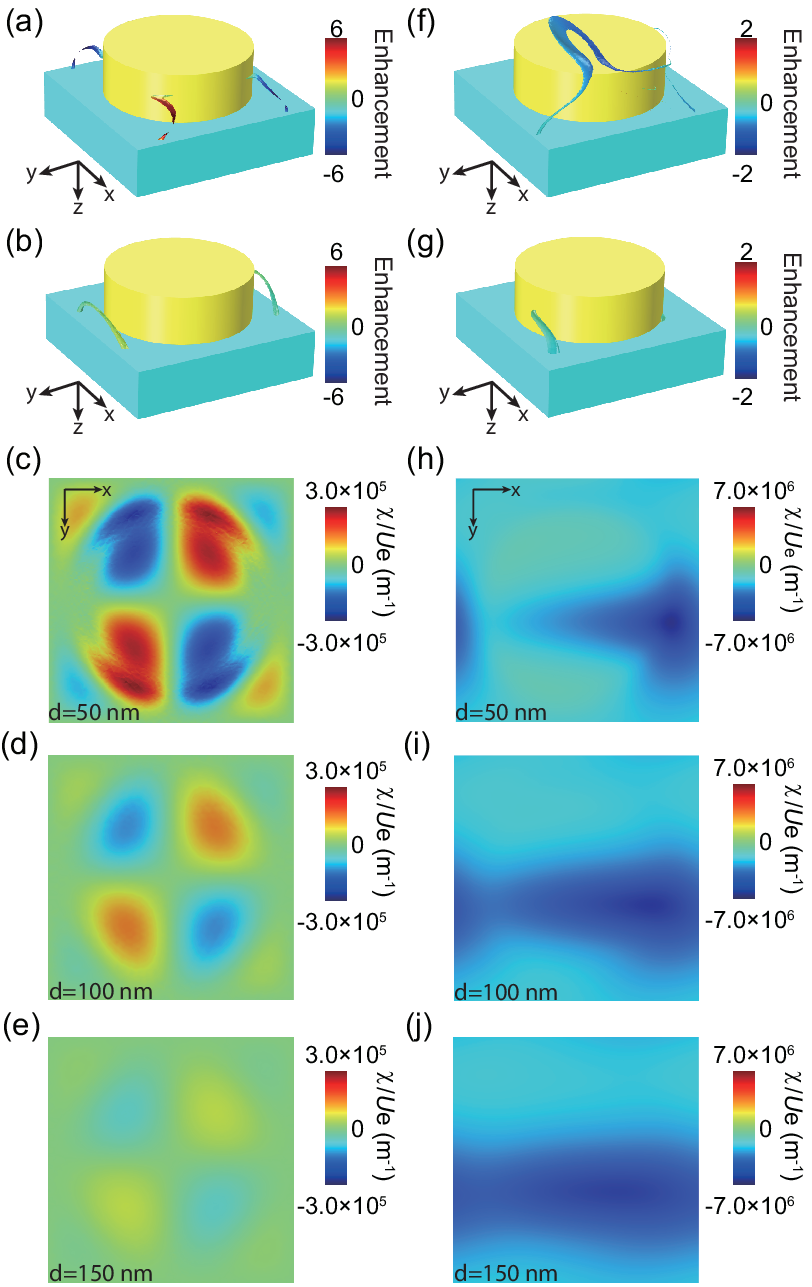}
        \caption{The (a) electric and (b) magnetic C lines induced in the unit cell under normal incidence. The corresponding normalized optical chirality is shown in (c)-(e) for three planes locating at \textit{d} = 50 nm, 100 nm, and 150 nm above the gold disk. The (f) electric and (g) magnetic C lines induced in the unit cell under oblique incidence ($\theta=60^{\circ}$,$\varphi=60^{\circ}$). The corresponding normalized optical chirality is shown in (h)-(j) for three planes locating at \textit{d} = 50 nm, 100 nm, and 150 nm above the gold disk.}
        \label{fig:2}
    \end{figure}
}

\newcommand{\FigThree}{
    \begin{figure}[t!]
        \centering
        \includegraphics[width=\linewidth]{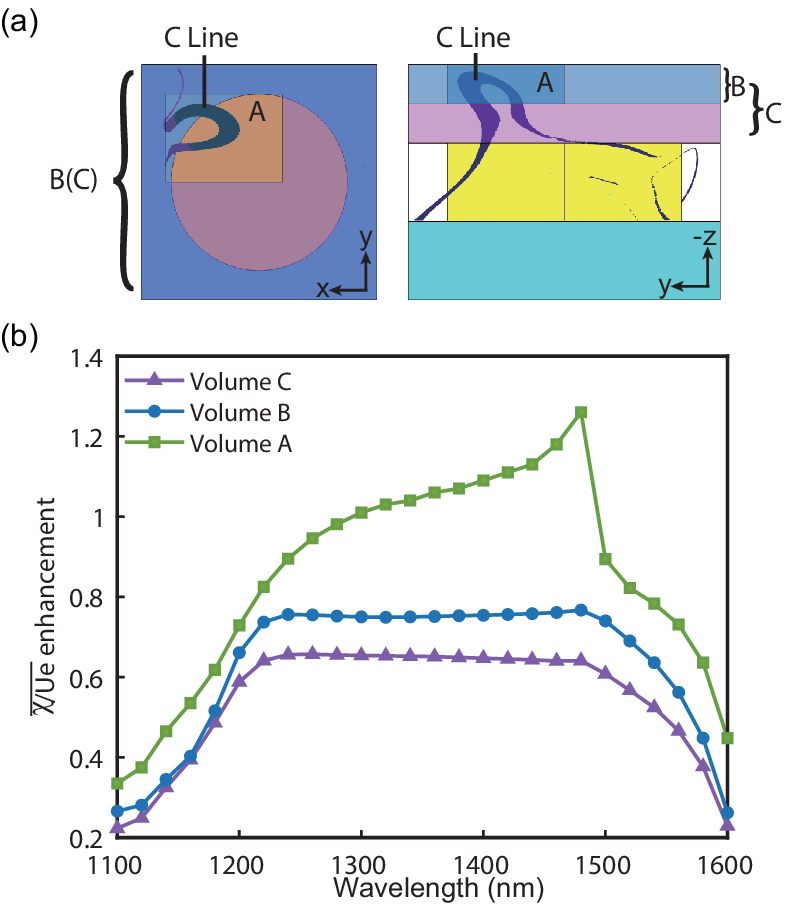}
        \caption{(a) Top and side views of the unit cell and the three different volumes A, B, C chosen for calculating the average of the normalized optical chirality. The smallest volume A (green) has a dimension of 400 nm $\times$ 300 nm $\times$ 100 nm and encloses a major part of the C line. The volume B (blue) has a dimension of $a \times a  \times $ 100 nm and encloses the volume A. The largest volume C (red) has a dimension of $a \times a  \times $ 200 nm and encloses the volume B. (b) Enhancement of the normalized optical chirality averaged over the three volumes compared to that of circularly polarized plane waves.}
        \label{fig:3}
    \end{figure}
}

\newcommand{\FigFour}{
    \begin{figure}[t!]
        \centering
        \includegraphics[width=\linewidth]{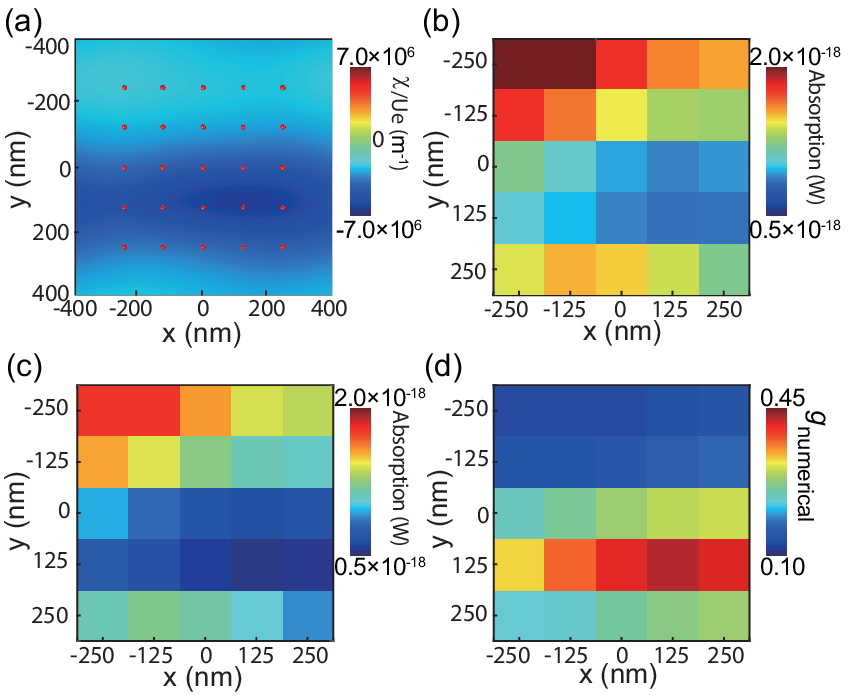}
        \caption{(a) The 25 sampling positions for calculating the absorption of the chiral particles. All the points locate on a plane that is 150 nm above top surface of the gold disk. The absorption of the (b) LH particle and (c) RH particle at the sampling positions. (d) The dissymmetry factor at the 25 sampling positions.}
        \label{fig:4}
    \end{figure}
}
\newcommand{\FigFive}{
    \begin{figure}[t!]
        \centering
        \includegraphics[width=\linewidth]{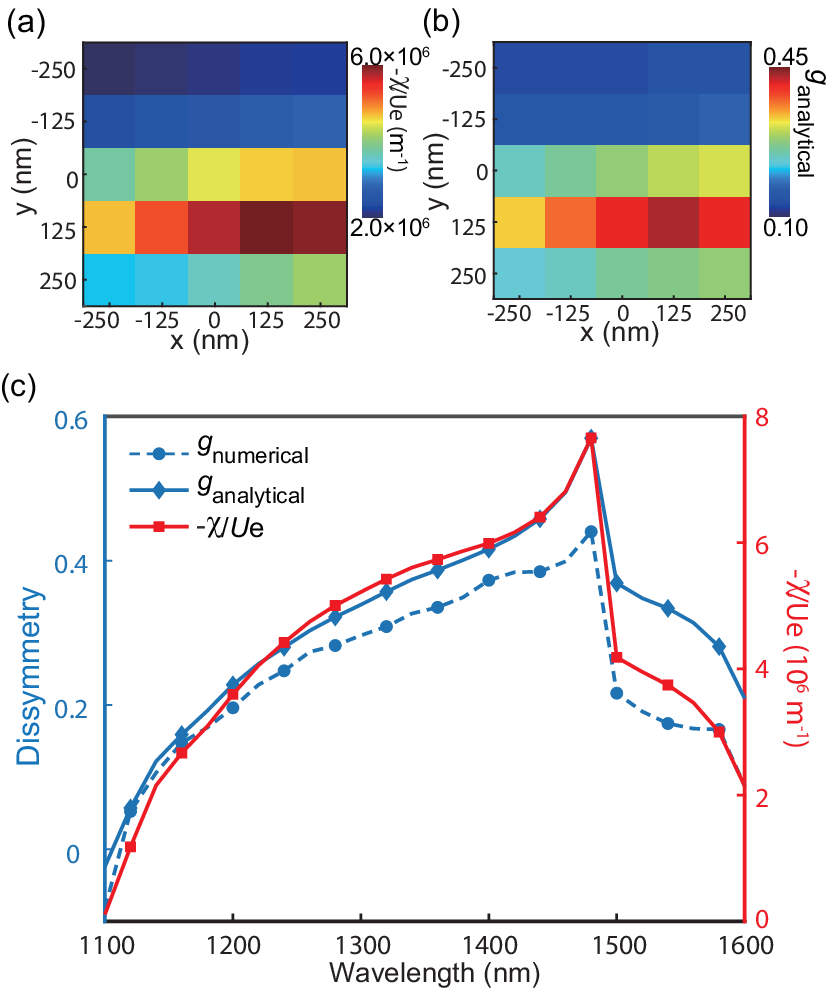}
        \caption{(a) Normalized optical chirality at the 25 sampling positions. (b) Analytical results of the $g$ factor at the 25 sampling positions. (c) The numerical and analytical $g$ factors and the normalized optical chirality at the position ($x=200$ nm, $y=100$ nm, $d=150$ nm) as a function of the wavelength.}
        \label{fig:5}
    \end{figure}
}

\begin{document}

\title{Broadband and large-area optical chirality generated by an achiral metasurface under achiral excitation}
\date{\today}

\author{Shiqi Jia}
\affiliation{Department of Physics, City University of Hong Kong, Tat Chee Avenue, Kowloon, Hong Kong, China}
\author{Tong Fu}
\affiliation{Department of Physics, City University of Hong Kong, Tat Chee Avenue, Kowloon, Hong Kong, China}
\author{Jie Peng}
\affiliation{Department of Physics, City University of Hong Kong, Tat Chee Avenue, Kowloon, Hong Kong, China}
\author{Shubo Wang}\email{shubwang@cityu.edu.hk}
\affiliation{Department of Physics, City University of Hong Kong, Tat Chee Avenue, Kowloon, Hong Kong, China}
\affiliation{City University of Hong Kong Shenzhen Research Institute, Shenzhen, Guangdong 518057, China}

\begin{abstract}
 Optical chirality plays an essential role in chiral light-matter interactions with broad applications in sensing and spectroscopy. Conventional methods of generating optical chirality usually employ chiral structures or chiral excitations. Here, we propose to use an achiral metasurface consisting of gold disk array excited by a linearly polarized light to generate optical chirality. Using full-wave numerical simulations, we show that the metasurface can give rise to large-area optical chirality of the same sign for the wavelength ranging from $1.2\ \si{\micro\metre}$ to $1.5\ \si{\micro\metre}$. The magnitude of the chirality is comparable to that of circularly polarized plane waves. The emergence of the optical chirality can be attributed to the asymmetric polarization singularity lines (C lines) in the near fields of the metasurface. We further explore the application of the proposed metasurface in chiral discriminations by simulating the  absorption of chiral helix particles immersed in the near fields, and demonstrate that the left-handed and right-handed helix particles give rise to different absorptions. The phenomenon can be understood using an analytical theory based on the dipole approximation, which predicts differential absorption quantitatively agrees with the numerical simulation results. Our study uncovers the subtle relationship between near-field optical chirality, polarization singularities, and symmetry. The results can find applications in optical sensing, chiral quantum optics, and optical manipulations of small particles. 
\end{abstract}
\maketitle

\section{\label{sec: I. Introduction}Introduction}
Chirality is a property of geometric symmetry with profound effects in various fields \cite{barronmolecular2004}. An object is chiral if it is non-superimposable with its mirror image. Chiral particles with opposite chirality (i.e., enantiomers) usually have different properties and can give rise to different functionalities \cite{wanglateral2014,zhang_optical_2015,chen_mechanical_2018,wo_optical_2020}. The discrimination of chiral particles is of critical importance to practical applications, such as in pharmaceutical and chemical industries \cite{eastgate_design_2017,suda_light-driven_2019,patel_pharmaceuticals_2019}. A widely used approach for chiral discriminations is circular dichroism (CD) spectroscopy, which is based on the enantiomers' preferential absorption of left-handed or right-handed circularly polarized light \cite{barronmolecular2004}. The CD signal is usually very weak due to the weak chiral light-matter interactions. Therefore, various approaches have been proposed to generate strong chiral optical fields to enhance chiral light-matter interactions, such as using standing waves \cite{tang_optical_2010}, vector beams with strong longitudinal components \cite{ye_enhancing_2021}, nanoresonators \cite{yoo_chiral_2015,mohammadi_dual_2021}, plasmonic nanostructures \cite{maoz_amplification_2013,hentschel_chiral_2017,graf_achiral_2019,lasaalonsoSurfaceEnhancedCircularDichroism2020a,garcia-guiradoEnhancedChiralSensing2020}, and metasurfaces \cite{mohammadiNanophotonicPlatformsEnhanced2018a,mohammadi_accessible_2019,gorkunovMetasurfacesMaximumChirality2020a,nguyen_large_2023,hong_nonlocal_2023}.

Recently, chiral optical fields generated by achiral structures have attracted a lot of attention since the achiral structures will not induce a chiral background signal \cite{wu_competition_2014}. Such chiral optical fields can be applied to enhance the CD signal and chiral sensing \cite{vazquez-guardado_superchiral_2018,solomon_enantiospecific_2019,feis_helicity-preserving_2020,czajkowskiLocalBulkCircular2022,ruiSurfaceEnhancedCircularDichroism2022}. To further reduce the complexity of the system, achiral structures excited by linearly polarized incident light (i.e., achiral light) have been employed to generate chiral optical fields \cite{liu_enhanced_2022}. It has been shown that C points\textemdash polarization singularities where the light is circularly polarized and the major axis of the polarization ellipse is ill-defined \cite{WaveStructureMonochromatic1987}\textemdash can give rise to chiral fields in simple achiral structures under linearly polarized plane wave excitation \cite{dennis_chapter_2009,peng_polarization_2021,jia_chiral_2022,peng_topological_2022}. However, the chiral optical fields in these systems are highly localized, and the optical chirality is usually nonuniform with its sign varying in different regions, which greatly impedes their applications in large-scale chiral light-matter interactions. In this work, we propose a simple achiral metasurface consisting of gold disk array to generate large-area chiral optical fields under the excitation of a linearly polarized light. By tuning the incident angle of the linearly polarized light, the metasurface can induce asymmetrical C lines (i.e., lines of C points). Importantly, the near fields of the metasurface carry optical chirality of the same sign in the whole unit cell. In addition, we study the absorption of chiral helix particles located inside the chiral fields and demonstrate the large absorption dissymmetry (i.e., CD effect) of the chiral particles with opposite handedness. We apply the dipole approximation to understand the relationship between the optical chirality and the absorption dissymmetry, which produces analytical results quantitatively agree with the numerical results. 

The paper is organized as follows. In Sec. II. A, we present the chiral fields and the C lines generated by the metasurface under the excitation of an obliquely incident linearly polarized plane wave. In Sec. II. B, we discuss the absorption properties of isotropic chiral helix particles in the chiral optical fields. To understand the phenomenon of absorption dissymmetry, we compare the numerical results with the analytical results based on the dipole approximation. We draw the conclusion in Sec. III.

\section{\label{sec: II. MST,SourceRep}	RESULTS AND DISCUSSION }
\subsection{\label{sec: A} Optical chirality and polarization singularities}
We consider a metasurface composed of periodic gold disks sitting on a silica ($\varepsilon_r=3.2$) substrate under the incidence of a linearly polarized plane wave with the electric field amplitude $\mathrm{E_0}=1$ V/m, as shown in Fig. \ref{fig:1}(a). The metasurface has a lattice constant $a = 800$ nm. The thickness of the silica substrate is $t=$ 200 nm. The gold disk has radius $r = 200$ nm and height $h = 200$ nm. The relative permittivity of gold is described by the Drude model $\varepsilon_\text{Au}=1-\omega_\text{p}^2\text{/}(\omega^2+i\omega\gamma)$, where $\omega_\text{p}=1.28\times10^{16}$  rad/s and $\gamma=7.10\times10^{13}$ rad/s \cite{olmon_optical_2012}. The incident wavevector $\mathbf{k}$ is in the $yoz-$plane and forms an angle of $\theta$ (i. e., incident angle) with respect to the $+z$ axis. The polarization angle $\phi$ is defined as the angle between the polarization vector and $yoz-$plane.
Under the excitation of the linearly polarized plane wave, the gold disks generate strong near fields. The chiral property of the near fields can be characterized by the normalized optical chirality $\chi/U_e$, where the optical chirality is defined as \cite{tang_optical_2010,vazquez-lozano_optical_2018}:
\EqnOne
$U_e=\frac{1}{4}\varepsilon_0|\mathbf{E}|^2$ is the electric field energy density, and $\varepsilon_0$ is permittivity of free space.

To understand the overall chirality of the near fields, we conduct full-wave numerical simulations using COMSOL Multiphysics and calculate the average of $\chi/U_e$ over the red-colored volume locating 100 nm above the gold disk in Fig. \ref{fig:1}(a), which has the dimensions of $a \times a \times$100 nm. The results, denoted as $\overline{\chi/U_e}$ , are shown in Fig. \ref{fig:1}(b) for different wavelengths. The blue line denotes $\overline{\chi/U_e}$ for the incident and polarization angles $\theta=\varphi=60$ degrees. We notice that the normalized optical chirality is negative in the considered wavelength range $\lambda\in[1.1\ \si{\micro\metre},1.6\  \si{\micro\metre}]$, and it varies slowly when the wavelength changes from $\lambda=1.2\ \si{\micro\metre}$ to $\lambda=1.5\ \si{\micro\metre}$. This indicates that the metasurface can generate broadband optical chirality in the near fields. For comparison, we also calculate $\overline{\chi/U_e}$ for normal incidence as denoted by the red line, which is zero at all wavelengths due to the mirror symmetry of the system. To further understand the dependence of optical chirality on the excitation, we calculate $\overline{\chi/U_e}$ at different incident angles and polarization angles. The results are shown in Fig. \ref{fig:1}(c) for the incident angle $\theta\in[5, 85]$ degrees and the polarization angle $\varphi\in[5, 85]$ degrees. We see that the absolute magnitude of $\overline{\chi/U_e}$ increases as $\theta$ and $\varphi$ increase, indicating that the optical chirality is attributed to the symmetry breaking under oblique incidence of the plane wave.
\FigOne
\FigTwo
\FigThree

To understand the mechanism for the emergence of the optical chirality, we determined the C lines in the unit cell under normal and oblique incidences. Figure \ref{fig:2}(a) shows the electric C lines under normal incidence at the wavelength $\lambda=1.4\ \si{\micro\metre}$. Two pairs of C lines grow from the side surface of the gold disk and extend into the substrate. These C lines correspond to the spatial locations where the electric field is circularly polarized. Thus, they carry optical chirality in general. The color of the C lines shows the enhancement of the normalized optical chirality compared to that of a circularly polarized plane wave. As seen, the optical chirality of the electric C lines is several times larger than that of a circularly polarized plane wave, and the sign of the optical chirality is mirror symmetric with respect to the \textit{xoz}- and \textit{yoz}- planes. Figure \ref{fig:2}(b) shows the C lines of the magnetic field, which are also mirror symmetric. We notice that the optical chirality of the magnetic C lines is much smaller than that of the electric C lines. This C-line configuration directly determines the distribution of optical chirality in the near fields. To understand this distribution, we calculate $\chi/U_e$ on different cutting planes at $d=50$ nm, $100$ nm, and $150$ nm above the metasurface, as shown in Fig. \ref{fig:2}(c), \ref{fig:2}(d), and \ref{fig:2}(e), respectively. As expected, the distribution of optical chirality is mirror symmetric, i.e., $\chi(x,y)=-\chi(-x,y)=-\chi(x,-y)=\chi(-x,-y)$. Therefore, although the absolute magnitude of the normalized optical chirality on the electric C lines is larger than that of a circularly polarized plane wave, the volume average $\overline{\chi/U_e}$ is zero. In addition, we notice that the absolute magnitude of the optical chirality rapidly decreases as the distance $d$ increases, which can be understood since the C lines mainly localize near the silica substrate, as shown in Fig. \ref{fig:2}(a). As a comparison, Fig. \ref{fig:2}(f) and \ref{fig:2}(g) show the electric and magnetic C lines, respectively, under the oblique incidence with $\theta=\varphi=60$ degrees at the wavelength $\lambda=1400$ nm. Since the oblique incidence breaks the mirror symmetry, the C lines do not emerge in pairs. Instead, there is an electric C line connecting the top surface of the disk and the substrate surface and a magnetic C line connecting the side surface of the disk and the substrate surface. Similar to the case of normal incidence, the optical chirality of the magnetic C line is much smaller than that of the electric C line. Figure \ref{fig:2}(h), \ref{fig:2}(i) and \ref{fig:2}(j) show the normalized optical chirality $\chi/U_e$ on three cutting planes at $d=50$ nm, $100$ nm, and $150$ nm above the top surface of gold disk, respectively. Notably, the optical chirality on the whole plane is negative, and its absolute magnitude increases as $d$ increases. Evidently, the distribution of optical chirality is directly decided by the C lines whose configurations depend on the symmetry of the system. We note that such asymmetric  C lines and optical chirality can also be generated at other incident and polarization angles. In addition, although the optical chirality is dominated by the electric C lines in the considered cases, the contribution of the magnetic C lines may not be negligible under other conditions (e.g, different frequency or excitation).

To further understand the relationship between the C lines and the optical chirality, we calculate $\overline{\chi/U_e}$ (i.e., volume average of the normalized optical chirality) for three different volumes highlighted in green, blue, and red in Fig. \ref{fig:3}(a) (labeled as A, B, and C). The smallest volume A (green) has a dimension of 400 nm $\times 300$ nm $\times 100$ nm and encloses a major part of the C line. The volume B (blue) has a dimension of $a \times a \times$100 nm and encloses the volume A. The volume C (red) has a dimension of $a \times a \times$200 nm and encloses the volume B. Figure \ref{fig:3}(b) shows the enhancement of $\overline{\chi/U_e}$ in the three regions compared to that of a circularly polarized plane wave. As seen, the chirality is the strongest in the volume A (denoted by the green line), which has a maximum value that is about 1.3 times the optical chirality of a circularly polarized plane wave. The $\overline{\chi/U_e}$ in the volume B (denoted by the blue line) is smaller than that of volume A, and the volume C has the smallest $\overline{\chi/U_e}$ (denoted by the purple line). These results demonstrate that the near-field chirality is indeed attributed to the emergence of the electric C lines. The field near the C line carries stronger chirality than the field far from the C line.

\subsection{\label{sec: B}	Absorption of chiral particles in the near fields}
We apply the near-field optical chirality generated by the metasurface to sense chiral particles through the CD effect. The efficiency of chiral sensing can be characterized by the dissymmetry in the absorption rate of the chiral particles, corresponding to the dissymmetry factor:
\EqnTwo
where $A^\text{L}$ and $A^\text{R}$ are the absorption rate of the left-handed (LH) or right-handed (RH) particles, respectively. In the numerical simulations, we use an isotropic chiral particle composed of metallic helices locating above the metasurface, as shown in Fig. \ref{fig:1}(a). The chiral particle contains six identical two-turn gold helices that are arranged in pairs in the three orthogonal directions. The composite particle is 100 nm long in each direction. The gold helix has a minor radius of 5 nm, a major radius of 10 nm, and its pitch is 15 nm. We simulated the absorption of the LH and RH chiral particles locating at $d = 150$ nm above the gold disk's upper surface. We consider a total of 25 sampling positions as marked by the red dots in Fig. \ref{fig:4}(a), where the color denotes the normalized optical chirality $\chi/U_e$ [same as the results in Fig. \ref{fig:2}(j)]. Figure \ref{fig:4}(b) and \ref{fig:4}(c) show the corresponding absorption rates of the LH and RH chiral particles locating at these sampling positions, respectively. As seen, the LH and RH particles have different absorptions. Figure \ref{fig:4}(d) shows the dissymmetry factor $g$ calculated using Eq. (\ref{eqn:2}), which agrees with the distribution of the normalized optical chirality in Fig. \ref{fig:4}(a). The dissymmetry factor reaches a maximum value of 0.45 at the spatial locations with maximum normalized optical chirality. 
\FigFour
To intuitively understand the above numerical results of absorption dissymmetry, we treat the subwavelength chiral particles as electric and magnetic dipoles and retrieve the polarizabilities based on scattering analysis. The chiral particle is approximately isotropic and can be characterized by the scalar electric polarizability $\alpha_e$, magnetic polarizability $\alpha_m$, and magnetoelectric polarizability $\kappa$.	The absorption of the chiral particles can be expressed as:
\EqnThree
where \textbf{p} and \textbf{m} are the electric dipole moment and magnetic dipole moment induced in the chiral particle, respectively. Using the dipole polarizabilities, we obtain:
\EqnFour
Substituting it into Eq. (\ref{eqn:3}), the absorption rate can be reduced to:
\EqnFive
where the `-' and `+' correspond to the LH and RH chiral particles, respectively; $\alpha_e^{\prime \prime}, \alpha_m^{\prime \prime}, \kappa^{\prime \prime}$ denote the imaginary parts of the polarizabilities. For anisotropic particles, the first two terms in Eq. (\ref{eqn:5}) also contribute to the absorption difference \cite{graf_achiral_2019,jia_chiral_2022}. For isotropic particles, the only contribution comes from the third term, and the corresponding absorption dissymmetry factor $g$ can be obtained as \cite{tang_optical_2010}:
\EqnSix
where we have assumed that the absorption due to $\alpha_m^{\prime \prime}$ is much smaller than the electric counterpart. Equation (\ref{eqn:6}) shows that the absorption dissymmetry is proportional to the normalized optical chirality $\chi/U_e$.

\FigFive

We apply Eq. (\ref{eqn:6}) to understand the relationship between the absorption dissymmetry and the optical chirality. Their comparison is shown in Fig. \ref{fig:5}(a) and \ref{fig:5}(b) for the 25 sampling positions. We notice the consistency between the two, confirming that the absorption dissymmetry is indeed attributed to the optical chirality. The analytical results in Fig. \ref{fig:5}(a) also agrees with the numerical results in Fig. \ref{fig:4}(d), demonstrating the validity of the analytical theory based on the dipole approximation. We further compare the numerical and analytical results of the dissymmetry factor at different wavelengths for the same sampling point at $x = 200$ nm, $y = 100$ nm, and $d = 150$ nm. The results are shown in Fig. \ref{fig:5}(c). As seen, the numerical and analytical results qualitatively agree with each other, and both are consistent with the normalized optical chirality. This further demonstrates the robustness of the analytical theory based on the dipole approximation.

\section{\label{sec: III. MST,SourceRep} CONCLUSION}
In conclusion, we propose a method to detect the chirality of small particles by using the chiral near field of an achiral metasurface excited by linearly polarized light at oblique incidence. We show that the metasurface can generate asymmetric C lines and give rise to large-area, broadband optical chirality of the same sign. We uncover the relationship between the optical chirality and the electric C lines. The optical chirality near the electric C lines can be larger than that of circularly polarized plane waves. We have applied the metasurface to detect chiral particles by simulating their absorption. Our analytical results based on the dipole approximation shows that the absorption of the chiral particles is attributed to the electric and magnetic dipoles induced in the chiral particles, and the absorption dissymmetry is proportional to the chirality of the particles and the normalized optical chirality of the field. 

Our study provides a convenient approach for broad-band chiral discrimination without using chiral excitations or chiral structures. The proposed approach uses a simple metasurface consisting of gold disks and capable of generating large-volume optical chirality of the same sign. Thus, it can be integrated into on-chip optical devices for large-area detection of chiral molecules. The results may also find applications in optical sensing, chiral quantum optics, and optical manipulations of small particles. 

\section{\label{sec: VII. Acknowledgements}Acknowledgements}
The work described in this paper was supported by the Research Grants Council of the Hong Kong Special Administrative Region, China (Project Nos. CityU 11306019, AoE/P-502/20, and C6013-18G).
\bibliography{references_mendeley}

\begin{thebibliography}{36}%
\makeatletter
\providecommand \@ifxundefined [1]{%
 \@ifx{#1\undefined}
}%
\providecommand \@ifnum [1]{%
 \ifnum #1\expandafter \@firstoftwo
 \else \expandafter \@secondoftwo
 \fi
}%
\providecommand \@ifx [1]{%
 \ifx #1\expandafter \@firstoftwo
 \else \expandafter \@secondoftwo
 \fi
}%
\providecommand \natexlab [1]{#1}%
\providecommand \enquote  [1]{``#1''}%
\providecommand \bibnamefont  [1]{#1}%
\providecommand \bibfnamefont [1]{#1}%
\providecommand \citenamefont [1]{#1}%
\providecommand \href@noop [0]{\@secondoftwo}%
\providecommand \href [0]{\begingroup \@sanitize@url \@href}%
\providecommand \@href[1]{\@@startlink{#1}\@@href}%
\providecommand \@@href[1]{\endgroup#1\@@endlink}%
\providecommand \@sanitize@url [0]{\catcode `\\12\catcode `\$12\catcode
  `\&12\catcode `\#12\catcode `\^12\catcode `\_12\catcode `\%12\relax}%
\providecommand \@@startlink[1]{}%
\providecommand \@@endlink[0]{}%
\providecommand \url  [0]{\begingroup\@sanitize@url \@url }%
\providecommand \@url [1]{\endgroup\@href {#1}{\urlprefix }}%
\providecommand \urlprefix  [0]{URL }%
\providecommand \Eprint [0]{\href }%
\providecommand \doibase [0]{https://doi.org/}%
\providecommand \selectlanguage [0]{\@gobble}%
\providecommand \bibinfo  [0]{\@secondoftwo}%
\providecommand \bibfield  [0]{\@secondoftwo}%
\providecommand \translation [1]{[#1]}%
\providecommand \BibitemOpen [0]{}%
\providecommand \bibitemStop [0]{}%
\providecommand \bibitemNoStop [0]{.\EOS\space}%
\providecommand \EOS [0]{\spacefactor3000\relax}%
\providecommand \BibitemShut  [1]{\csname bibitem#1\endcsname}%
\let\auto@bib@innerbib\@empty
\bibitem [{\citenamefont {Barron}(2004)}]{barronmolecular2004}%
  \BibitemOpen
  \bibfield  {author} {\bibinfo {author} {\bibfnamefont {L.~D.}\ \bibnamefont
  {Barron}},\ }\href {https://doi.org/10.1017/CBO9780511535468} {\emph
  {\bibinfo {title} {Molecular Light Scattering and Optical Activity}}},\
  \bibinfo {edition} {2nd}\ ed.\ (\bibinfo  {publisher} {Cambridge University
  Press},\ \bibinfo {year} {2004})\BibitemShut {NoStop}%
\bibitem [{\citenamefont {Wang}\ and\ \citenamefont
  {Chan}(2014)}]{wanglateral2014}%
  \BibitemOpen
  \bibfield  {author} {\bibinfo {author} {\bibfnamefont {S.~B.}\ \bibnamefont
  {Wang}}\ and\ \bibinfo {author} {\bibfnamefont {C.~T.}\ \bibnamefont
  {Chan}},\ }\href {https://doi.org/10.1038/ncomms4307} {\bibfield  {journal}
  {\bibinfo  {journal} {Nat. Commun.}\ }\textbf {\bibinfo {volume} {5}},\
  \bibinfo {pages} {3307} (\bibinfo {year} {2014})}\BibitemShut {NoStop}%
\bibitem [{\citenamefont {Zhang}\ \emph {et~al.}(2015)\citenamefont {Zhang},
  \citenamefont {Wang}, \citenamefont {Lin}, \citenamefont {Sun},\ and\
  \citenamefont {Chan}}]{zhang_optical_2015}%
  \BibitemOpen
  \bibfield  {author} {\bibinfo {author} {\bibfnamefont {X.-L.}\ \bibnamefont
  {Zhang}}, \bibinfo {author} {\bibfnamefont {S.~B.}\ \bibnamefont {Wang}},
  \bibinfo {author} {\bibfnamefont {Z.}~\bibnamefont {Lin}}, \bibinfo {author}
  {\bibfnamefont {H.-B.}\ \bibnamefont {Sun}},\ and\ \bibinfo {author}
  {\bibfnamefont {C.~T.}\ \bibnamefont {Chan}},\ }\href
  {https://doi.org/10.1103/PhysRevA.92.043804} {\bibfield  {journal} {\bibinfo
  {journal} {Phys. Rev. A}\ }\textbf {\bibinfo {volume} {92}},\ \bibinfo
  {pages} {043804} (\bibinfo {year} {2015})}\BibitemShut {NoStop}%
\bibitem [{\citenamefont {Chen}\ \emph {et~al.}(2018)\citenamefont {Chen},
  \citenamefont {Wang}, \citenamefont {Li},\ and\ \citenamefont
  {Ng}}]{chen_mechanical_2018}%
  \BibitemOpen
  \bibfield  {author} {\bibinfo {author} {\bibfnamefont {J.}~\bibnamefont
  {Chen}}, \bibinfo {author} {\bibfnamefont {S.}~\bibnamefont {Wang}}, \bibinfo
  {author} {\bibfnamefont {X.}~\bibnamefont {Li}},\ and\ \bibinfo {author}
  {\bibfnamefont {J.}~\bibnamefont {Ng}},\ }\href
  {https://doi.org/10.1364/OE.26.027694} {\bibfield  {journal} {\bibinfo
  {journal} {Opt. Express, {OE}}\ }\textbf {\bibinfo {volume} {26}},\ \bibinfo
  {pages} {27694} (\bibinfo {year} {2018})}\BibitemShut {NoStop}%
\bibitem [{\citenamefont {Wo}\ \emph {et~al.}(2020)\citenamefont {Wo},
  \citenamefont {Peng}, \citenamefont {Prasad}, \citenamefont {Shi},
  \citenamefont {Li},\ and\ \citenamefont {Wang}}]{wo_optical_2020}%
  \BibitemOpen
  \bibfield  {author} {\bibinfo {author} {\bibfnamefont {K.~J.}\ \bibnamefont
  {Wo}}, \bibinfo {author} {\bibfnamefont {J.}~\bibnamefont {Peng}}, \bibinfo
  {author} {\bibfnamefont {M.~K.}\ \bibnamefont {Prasad}}, \bibinfo {author}
  {\bibfnamefont {Y.}~\bibnamefont {Shi}}, \bibinfo {author} {\bibfnamefont
  {J.}~\bibnamefont {Li}},\ and\ \bibinfo {author} {\bibfnamefont
  {S.}~\bibnamefont {Wang}},\ }\href
  {https://doi.org/10.1103/PhysRevA.102.043526} {\bibfield  {journal} {\bibinfo
   {journal} {Phys. Rev. A}\ }\textbf {\bibinfo {volume} {102}},\ \bibinfo
  {pages} {043526} (\bibinfo {year} {2020})}\BibitemShut {NoStop}%
\bibitem [{\citenamefont {Eastgate}\ \emph {et~al.}(2017)\citenamefont
  {Eastgate}, \citenamefont {Schmidt},\ and\ \citenamefont
  {Fandrick}}]{eastgate_design_2017}%
  \BibitemOpen
  \bibfield  {author} {\bibinfo {author} {\bibfnamefont {M.~D.}\ \bibnamefont
  {Eastgate}}, \bibinfo {author} {\bibfnamefont {M.~A.}\ \bibnamefont
  {Schmidt}},\ and\ \bibinfo {author} {\bibfnamefont {K.~R.}\ \bibnamefont
  {Fandrick}},\ }\href {https://doi.org/10.1038/s41570-017-0016} {\bibfield
  {journal} {\bibinfo  {journal} {Nat Rev Chem}\ }\textbf {\bibinfo {volume}
  {1}},\ \bibinfo {pages} {1} (\bibinfo {year} {2017})}\BibitemShut {NoStop}%
\bibitem [{\citenamefont {Suda}\ \emph {et~al.}(2019)\citenamefont {Suda},
  \citenamefont {Thathong}, \citenamefont {Promarak}, \citenamefont {Kojima},
  \citenamefont {Nakamura}, \citenamefont {Shiraogawa}, \citenamefont {Ehara},\
  and\ \citenamefont {Yamamoto}}]{suda_light-driven_2019}%
  \BibitemOpen
  \bibfield  {author} {\bibinfo {author} {\bibfnamefont {M.}~\bibnamefont
  {Suda}}, \bibinfo {author} {\bibfnamefont {Y.}~\bibnamefont {Thathong}},
  \bibinfo {author} {\bibfnamefont {V.}~\bibnamefont {Promarak}}, \bibinfo
  {author} {\bibfnamefont {H.}~\bibnamefont {Kojima}}, \bibinfo {author}
  {\bibfnamefont {M.}~\bibnamefont {Nakamura}}, \bibinfo {author}
  {\bibfnamefont {T.}~\bibnamefont {Shiraogawa}}, \bibinfo {author}
  {\bibfnamefont {M.}~\bibnamefont {Ehara}},\ and\ \bibinfo {author}
  {\bibfnamefont {H.~M.}\ \bibnamefont {Yamamoto}},\ }\href
  {https://doi.org/10.1038/s41467-019-10423-6} {\bibfield  {journal} {\bibinfo
  {journal} {Nat Commun}\ }\textbf {\bibinfo {volume} {10}},\ \bibinfo {pages}
  {2455} (\bibinfo {year} {2019})}\BibitemShut {NoStop}%
\bibitem [{\citenamefont {Patel}\ \emph {et~al.}(2019)\citenamefont {Patel},
  \citenamefont {Kumar}, \citenamefont {Kishor}, \citenamefont {Mlsna},
  \citenamefont {Pittman},\ and\ \citenamefont
  {Mohan}}]{patel_pharmaceuticals_2019}%
  \BibitemOpen
  \bibfield  {author} {\bibinfo {author} {\bibfnamefont {M.}~\bibnamefont
  {Patel}}, \bibinfo {author} {\bibfnamefont {R.}~\bibnamefont {Kumar}},
  \bibinfo {author} {\bibfnamefont {K.}~\bibnamefont {Kishor}}, \bibinfo
  {author} {\bibfnamefont {T.}~\bibnamefont {Mlsna}}, \bibinfo {author}
  {\bibfnamefont {C.~U.}\ \bibnamefont {Pittman}},\ and\ \bibinfo {author}
  {\bibfnamefont {D.}~\bibnamefont {Mohan}},\ }\href
  {https://doi.org/10.1021/acs.chemrev.8b00299} {\bibfield  {journal} {\bibinfo
   {journal} {Chem. Rev.}\ }\textbf {\bibinfo {volume} {119}},\ \bibinfo
  {pages} {3510} (\bibinfo {year} {2019})}\BibitemShut {NoStop}%
\bibitem [{\citenamefont {Tang}\ and\ \citenamefont
  {Cohen}(2010)}]{tang_optical_2010}%
  \BibitemOpen
  \bibfield  {author} {\bibinfo {author} {\bibfnamefont {Y.}~\bibnamefont
  {Tang}}\ and\ \bibinfo {author} {\bibfnamefont {A.~E.}\ \bibnamefont
  {Cohen}},\ }\href {https://doi.org/10.1103/PhysRevLett.104.163901} {\bibfield
   {journal} {\bibinfo  {journal} {Phys. Rev. Lett.}\ }\textbf {\bibinfo
  {volume} {104}},\ \bibinfo {pages} {163901} (\bibinfo {year}
  {2010})}\BibitemShut {NoStop}%
\bibitem [{\citenamefont {Ye}\ \emph {et~al.}(2021)\citenamefont {Ye},
  \citenamefont {Yang}, \citenamefont {Zheng},\ and\ \citenamefont
  {Mukamel}}]{ye_enhancing_2021}%
  \BibitemOpen
  \bibfield  {author} {\bibinfo {author} {\bibfnamefont {L.}~\bibnamefont
  {Ye}}, \bibinfo {author} {\bibfnamefont {L.}~\bibnamefont {Yang}}, \bibinfo
  {author} {\bibfnamefont {X.}~\bibnamefont {Zheng}},\ and\ \bibinfo {author}
  {\bibfnamefont {S.}~\bibnamefont {Mukamel}},\ }\href
  {https://doi.org/10.1103/PhysRevLett.126.123001} {\bibfield  {journal}
  {\bibinfo  {journal} {Phys. Rev. Lett.}\ }\textbf {\bibinfo {volume} {126}},\
  \bibinfo {pages} {123001} (\bibinfo {year} {2021})}\BibitemShut {NoStop}%
\bibitem [{\citenamefont {Yoo}\ and\ \citenamefont
  {Park}(2015)}]{yoo_chiral_2015}%
  \BibitemOpen
  \bibfield  {author} {\bibinfo {author} {\bibfnamefont {S.}~\bibnamefont
  {Yoo}}\ and\ \bibinfo {author} {\bibfnamefont {Q.-H.}\ \bibnamefont {Park}},\
  }\href {https://doi.org/10.1103/PhysRevLett.114.203003} {\bibfield  {journal}
  {\bibinfo  {journal} {Phys. Rev. Lett.}\ }\textbf {\bibinfo {volume} {114}},\
  \bibinfo {pages} {203003} (\bibinfo {year} {2015})}\BibitemShut {NoStop}%
\bibitem [{\citenamefont {Mohammadi}\ \emph {et~al.}(2021)\citenamefont
  {Mohammadi}, \citenamefont {Tittl}, \citenamefont {Tsakmakidis},
  \citenamefont {Raziman},\ and\ \citenamefont {Curto}}]{mohammadi_dual_2021}%
  \BibitemOpen
  \bibfield  {author} {\bibinfo {author} {\bibfnamefont {E.}~\bibnamefont
  {Mohammadi}}, \bibinfo {author} {\bibfnamefont {A.}~\bibnamefont {Tittl}},
  \bibinfo {author} {\bibfnamefont {K.~L.}\ \bibnamefont {Tsakmakidis}},
  \bibinfo {author} {\bibfnamefont {T.~V.}\ \bibnamefont {Raziman}},\ and\
  \bibinfo {author} {\bibfnamefont {A.~G.}\ \bibnamefont {Curto}},\ }\href
  {https://doi.org/10.1021/acsphotonics.1c00311} {\bibfield  {journal}
  {\bibinfo  {journal} {{ACS} Photonics}\ }\textbf {\bibinfo {volume} {8}},\
  \bibinfo {pages} {1754} (\bibinfo {year} {2021})}\BibitemShut {NoStop}%
\bibitem [{\citenamefont {Maoz}\ \emph {et~al.}(2013)\citenamefont {Maoz},
  \citenamefont {Chaikin}, \citenamefont {Tesler}, \citenamefont {Bar~Elli},
  \citenamefont {Fan}, \citenamefont {Govorov},\ and\ \citenamefont
  {Markovich}}]{maoz_amplification_2013}%
  \BibitemOpen
  \bibfield  {author} {\bibinfo {author} {\bibfnamefont {B.~M.}\ \bibnamefont
  {Maoz}}, \bibinfo {author} {\bibfnamefont {Y.}~\bibnamefont {Chaikin}},
  \bibinfo {author} {\bibfnamefont {A.~B.}\ \bibnamefont {Tesler}}, \bibinfo
  {author} {\bibfnamefont {O.}~\bibnamefont {Bar~Elli}}, \bibinfo {author}
  {\bibfnamefont {Z.}~\bibnamefont {Fan}}, \bibinfo {author} {\bibfnamefont
  {A.~O.}\ \bibnamefont {Govorov}},\ and\ \bibinfo {author} {\bibfnamefont
  {G.}~\bibnamefont {Markovich}},\ }\href {https://doi.org/10.1021/nl304638a}
  {\bibfield  {journal} {\bibinfo  {journal} {Nano Lett.}\ }\textbf {\bibinfo
  {volume} {13}},\ \bibinfo {pages} {1203} (\bibinfo {year}
  {2013})}\BibitemShut {NoStop}%
\bibitem [{\citenamefont {Hentschel}\ \emph {et~al.}(2017)\citenamefont
  {Hentschel}, \citenamefont {Schäferling}, \citenamefont {Duan},
  \citenamefont {Giessen},\ and\ \citenamefont {Liu}}]{hentschel_chiral_2017}%
  \BibitemOpen
  \bibfield  {author} {\bibinfo {author} {\bibfnamefont {M.}~\bibnamefont
  {Hentschel}}, \bibinfo {author} {\bibfnamefont {M.}~\bibnamefont
  {Schäferling}}, \bibinfo {author} {\bibfnamefont {X.}~\bibnamefont {Duan}},
  \bibinfo {author} {\bibfnamefont {H.}~\bibnamefont {Giessen}},\ and\ \bibinfo
  {author} {\bibfnamefont {N.}~\bibnamefont {Liu}},\ }\href
  {https://doi.org/10.1126/sciadv.1602735} {\bibfield  {journal} {\bibinfo
  {journal} {Sci. Adv.}\ }\textbf {\bibinfo {volume} {3}},\ \bibinfo {pages}
  {e1602735} (\bibinfo {year} {2017})}\BibitemShut {NoStop}%
\bibitem [{\citenamefont {Graf}\ \emph {et~al.}(2019)\citenamefont {Graf},
  \citenamefont {Feis}, \citenamefont {Garcia-Santiago}, \citenamefont
  {Wegener}, \citenamefont {Rockstuhl},\ and\ \citenamefont
  {Fernandez-Corbaton}}]{graf_achiral_2019}%
  \BibitemOpen
  \bibfield  {author} {\bibinfo {author} {\bibfnamefont {F.}~\bibnamefont
  {Graf}}, \bibinfo {author} {\bibfnamefont {J.}~\bibnamefont {Feis}}, \bibinfo
  {author} {\bibfnamefont {X.}~\bibnamefont {Garcia-Santiago}}, \bibinfo
  {author} {\bibfnamefont {M.}~\bibnamefont {Wegener}}, \bibinfo {author}
  {\bibfnamefont {C.}~\bibnamefont {Rockstuhl}},\ and\ \bibinfo {author}
  {\bibfnamefont {I.}~\bibnamefont {Fernandez-Corbaton}},\ }\href
  {https://doi.org/10.1021/acsphotonics.8b01454} {\bibfield  {journal}
  {\bibinfo  {journal} {ACS Photonics}\ }\textbf {\bibinfo {volume} {6}},\
  \bibinfo {pages} {482} (\bibinfo {year} {2019})}\BibitemShut {NoStop}%
\bibitem [{\citenamefont {Lasa-Alonso}\ \emph {et~al.}(2020)\citenamefont
  {Lasa-Alonso}, \citenamefont {Abujetas}, \citenamefont {Nodar}, \citenamefont
  {Dionne}, \citenamefont {Sáenz}, \citenamefont {Molina-Terriza},
  \citenamefont {Aizpurua},\ and\ \citenamefont
  {García-Etxarri}}]{lasaalonsoSurfaceEnhancedCircularDichroism2020a}%
  \BibitemOpen
  \bibfield  {author} {\bibinfo {author} {\bibfnamefont {J.}~\bibnamefont
  {Lasa-Alonso}}, \bibinfo {author} {\bibfnamefont {D.~R.}\ \bibnamefont
  {Abujetas}}, \bibinfo {author} {\bibfnamefont {A.}~\bibnamefont {Nodar}},
  \bibinfo {author} {\bibfnamefont {J.~A.}\ \bibnamefont {Dionne}}, \bibinfo
  {author} {\bibfnamefont {J.~J.}\ \bibnamefont {Sáenz}}, \bibinfo {author}
  {\bibfnamefont {G.}~\bibnamefont {Molina-Terriza}}, \bibinfo {author}
  {\bibfnamefont {J.}~\bibnamefont {Aizpurua}},\ and\ \bibinfo {author}
  {\bibfnamefont {A.}~\bibnamefont {García-Etxarri}},\ }\href
  {https://doi.org/10.1021/acsphotonics.0c00611} {\bibfield  {journal}
  {\bibinfo  {journal} {ACS Photonics}\ }\textbf {\bibinfo {volume} {7}},\
  \bibinfo {pages} {2978} (\bibinfo {year} {2020})}\BibitemShut {NoStop}%
\bibitem [{\citenamefont {García-Guirado}\ \emph {et~al.}(2020)\citenamefont
  {García-Guirado}, \citenamefont {Svedendahl}, \citenamefont {Puigdollers},\
  and\ \citenamefont {Quidant}}]{garcia-guiradoEnhancedChiralSensing2020}%
  \BibitemOpen
  \bibfield  {author} {\bibinfo {author} {\bibfnamefont {J.}~\bibnamefont
  {García-Guirado}}, \bibinfo {author} {\bibfnamefont {M.}~\bibnamefont
  {Svedendahl}}, \bibinfo {author} {\bibfnamefont {J.}~\bibnamefont
  {Puigdollers}},\ and\ \bibinfo {author} {\bibfnamefont {R.}~\bibnamefont
  {Quidant}},\ }\href {https://doi.org/10.1021/acs.nanolett.9b04334} {\bibfield
   {journal} {\bibinfo  {journal} {Nano Lett.}\ }\textbf {\bibinfo {volume}
  {20}},\ \bibinfo {pages} {585} (\bibinfo {year} {2020})}\BibitemShut
  {NoStop}%
\bibitem [{\citenamefont {Mohammadi}\ \emph {et~al.}(2018)\citenamefont
  {Mohammadi}, \citenamefont {Tsakmakidis}, \citenamefont {Askarpour},
  \citenamefont {Dehkhoda}, \citenamefont {Tavakoli},\ and\ \citenamefont
  {Altug}}]{mohammadiNanophotonicPlatformsEnhanced2018a}%
  \BibitemOpen
  \bibfield  {author} {\bibinfo {author} {\bibfnamefont {E.}~\bibnamefont
  {Mohammadi}}, \bibinfo {author} {\bibfnamefont {K.~L.}\ \bibnamefont
  {Tsakmakidis}}, \bibinfo {author} {\bibfnamefont {A.~N.}\ \bibnamefont
  {Askarpour}}, \bibinfo {author} {\bibfnamefont {P.}~\bibnamefont {Dehkhoda}},
  \bibinfo {author} {\bibfnamefont {A.}~\bibnamefont {Tavakoli}},\ and\
  \bibinfo {author} {\bibfnamefont {H.}~\bibnamefont {Altug}},\ }\href
  {https://doi.org/10.1021/acsphotonics.8b00270} {\bibfield  {journal}
  {\bibinfo  {journal} {ACS Photonics}\ }\textbf {\bibinfo {volume} {5}},\
  \bibinfo {pages} {2669} (\bibinfo {year} {2018})}\BibitemShut {NoStop}%
\bibitem [{\citenamefont {Mohammadi}\ \emph {et~al.}(2019)\citenamefont
  {Mohammadi}, \citenamefont {Tavakoli}, \citenamefont {Dehkhoda},
  \citenamefont {Jahani}, \citenamefont {Tsakmakidis}, \citenamefont {Tittl},\
  and\ \citenamefont {Altug}}]{mohammadi_accessible_2019}%
  \BibitemOpen
  \bibfield  {author} {\bibinfo {author} {\bibfnamefont {E.}~\bibnamefont
  {Mohammadi}}, \bibinfo {author} {\bibfnamefont {A.}~\bibnamefont {Tavakoli}},
  \bibinfo {author} {\bibfnamefont {P.}~\bibnamefont {Dehkhoda}}, \bibinfo
  {author} {\bibfnamefont {Y.}~\bibnamefont {Jahani}}, \bibinfo {author}
  {\bibfnamefont {K.~L.}\ \bibnamefont {Tsakmakidis}}, \bibinfo {author}
  {\bibfnamefont {A.}~\bibnamefont {Tittl}},\ and\ \bibinfo {author}
  {\bibfnamefont {H.}~\bibnamefont {Altug}},\ }\href
  {https://doi.org/10.1021/acsphotonics.8b01767} {\bibfield  {journal}
  {\bibinfo  {journal} {{ACS} Photonics}\ }\textbf {\bibinfo {volume} {6}},\
  \bibinfo {pages} {1939} (\bibinfo {year} {2019})}\BibitemShut {NoStop}%
\bibitem [{\citenamefont {Gorkunov}\ \emph {et~al.}(2020)\citenamefont
  {Gorkunov}, \citenamefont {Antonov},\ and\ \citenamefont
  {Kivshar}}]{gorkunovMetasurfacesMaximumChirality2020a}%
  \BibitemOpen
  \bibfield  {author} {\bibinfo {author} {\bibfnamefont {M.~V.}\ \bibnamefont
  {Gorkunov}}, \bibinfo {author} {\bibfnamefont {A.~A.}\ \bibnamefont
  {Antonov}},\ and\ \bibinfo {author} {\bibfnamefont {Y.~S.}\ \bibnamefont
  {Kivshar}},\ }\href {https://doi.org/10.1103/PhysRevLett.125.093903}
  {\bibfield  {journal} {\bibinfo  {journal} {Phys. Rev. Lett.}\ }\textbf
  {\bibinfo {volume} {125}},\ \bibinfo {pages} {093903} (\bibinfo {year}
  {2020})}\BibitemShut {NoStop}%
\bibitem [{\citenamefont {Nguyen}\ \emph {et~al.}(2023)\citenamefont {Nguyen},
  \citenamefont {Hugonin}, \citenamefont {Coutrot}, \citenamefont
  {Garcia-Caurel}, \citenamefont {Vest},\ and\ \citenamefont
  {Greffet}}]{nguyen_large_2023}%
  \BibitemOpen
  \bibfield  {author} {\bibinfo {author} {\bibfnamefont {A.}~\bibnamefont
  {Nguyen}}, \bibinfo {author} {\bibfnamefont {J.-P.}\ \bibnamefont {Hugonin}},
  \bibinfo {author} {\bibfnamefont {A.-L.}\ \bibnamefont {Coutrot}}, \bibinfo
  {author} {\bibfnamefont {E.}~\bibnamefont {Garcia-Caurel}}, \bibinfo {author}
  {\bibfnamefont {B.}~\bibnamefont {Vest}},\ and\ \bibinfo {author}
  {\bibfnamefont {J.-J.}\ \bibnamefont {Greffet}},\ }\href
  {https://doi.org/10.1364/OPTICA.480292} {\bibfield  {journal} {\bibinfo
  {journal} {Optica}\ }\textbf {\bibinfo {volume} {10}},\ \bibinfo {pages}
  {232} (\bibinfo {year} {2023})}\BibitemShut {NoStop}%
\bibitem [{\citenamefont {Hong}\ \emph {et~al.}(2023)\citenamefont {Hong},
  \citenamefont {van~de Groep}, \citenamefont {Lee}, \citenamefont {Kim},
  \citenamefont {Lalanne}, \citenamefont {Kik},\ and\ \citenamefont
  {Brongersma}}]{hong_nonlocal_2023}%
  \BibitemOpen
  \bibfield  {author} {\bibinfo {author} {\bibfnamefont {J.}~\bibnamefont
  {Hong}}, \bibinfo {author} {\bibfnamefont {J.}~\bibnamefont {van~de Groep}},
  \bibinfo {author} {\bibfnamefont {N.}~\bibnamefont {Lee}}, \bibinfo {author}
  {\bibfnamefont {S.~J.}\ \bibnamefont {Kim}}, \bibinfo {author} {\bibfnamefont
  {P.}~\bibnamefont {Lalanne}}, \bibinfo {author} {\bibfnamefont {P.~G.}\
  \bibnamefont {Kik}},\ and\ \bibinfo {author} {\bibfnamefont {M.~L.}\
  \bibnamefont {Brongersma}},\ }\href {https://doi.org/10.1364/OPTICA.468252}
  {\bibfield  {journal} {\bibinfo  {journal} {Optica}\ }\textbf {\bibinfo
  {volume} {10}},\ \bibinfo {pages} {134} (\bibinfo {year} {2023})}\BibitemShut
  {NoStop}%
\bibitem [{\citenamefont {Wu}\ \emph {et~al.}(2014)\citenamefont {Wu},
  \citenamefont {Ren}, \citenamefont {Wang},\ and\ \citenamefont
  {Zhang}}]{wu_competition_2014}%
  \BibitemOpen
  \bibfield  {author} {\bibinfo {author} {\bibfnamefont {T.}~\bibnamefont
  {Wu}}, \bibinfo {author} {\bibfnamefont {J.}~\bibnamefont {Ren}}, \bibinfo
  {author} {\bibfnamefont {R.}~\bibnamefont {Wang}},\ and\ \bibinfo {author}
  {\bibfnamefont {X.}~\bibnamefont {Zhang}},\ }\href
  {https://doi.org/10.1021/jp505290v} {\bibfield  {journal} {\bibinfo
  {journal} {J. Phys. Chem. C}\ }\textbf {\bibinfo {volume} {118}},\ \bibinfo
  {pages} {20529} (\bibinfo {year} {2014})}\BibitemShut {NoStop}%
\bibitem [{\citenamefont {Vázquez-Guardado}\ and\ \citenamefont
  {Chanda}(2018)}]{vazquez-guardado_superchiral_2018}%
  \BibitemOpen
  \bibfield  {author} {\bibinfo {author} {\bibfnamefont {A.}~\bibnamefont
  {Vázquez-Guardado}}\ and\ \bibinfo {author} {\bibfnamefont {D.}~\bibnamefont
  {Chanda}},\ }\href {https://doi.org/10.1103/PhysRevLett.120.137601}
  {\bibfield  {journal} {\bibinfo  {journal} {Phys. Rev. Lett.}\ }\textbf
  {\bibinfo {volume} {120}},\ \bibinfo {pages} {137601} (\bibinfo {year}
  {2018})}\BibitemShut {NoStop}%
\bibitem [{\citenamefont {Solomon}\ \emph {et~al.}(2019)\citenamefont
  {Solomon}, \citenamefont {Hu}, \citenamefont {Lawrence}, \citenamefont
  {García-Etxarri},\ and\ \citenamefont
  {Dionne}}]{solomon_enantiospecific_2019}%
  \BibitemOpen
  \bibfield  {author} {\bibinfo {author} {\bibfnamefont {M.~L.}\ \bibnamefont
  {Solomon}}, \bibinfo {author} {\bibfnamefont {J.}~\bibnamefont {Hu}},
  \bibinfo {author} {\bibfnamefont {M.}~\bibnamefont {Lawrence}}, \bibinfo
  {author} {\bibfnamefont {A.}~\bibnamefont {García-Etxarri}},\ and\ \bibinfo
  {author} {\bibfnamefont {J.~A.}\ \bibnamefont {Dionne}},\ }\href
  {https://doi.org/10.1021/acsphotonics.8b01365} {\bibfield  {journal}
  {\bibinfo  {journal} {{ACS} Photonics}\ }\textbf {\bibinfo {volume} {6}},\
  \bibinfo {pages} {43} (\bibinfo {year} {2019})}\BibitemShut {NoStop}%
\bibitem [{\citenamefont {Feis}\ \emph {et~al.}(2020)\citenamefont {Feis},
  \citenamefont {Beutel}, \citenamefont {Köpfler}, \citenamefont
  {Garcia-Santiago}, \citenamefont {Rockstuhl}, \citenamefont {Wegener},\ and\
  \citenamefont {Fernandez-Corbaton}}]{feis_helicity-preserving_2020}%
  \BibitemOpen
  \bibfield  {author} {\bibinfo {author} {\bibfnamefont {J.}~\bibnamefont
  {Feis}}, \bibinfo {author} {\bibfnamefont {D.}~\bibnamefont {Beutel}},
  \bibinfo {author} {\bibfnamefont {J.}~\bibnamefont {Köpfler}}, \bibinfo
  {author} {\bibfnamefont {X.}~\bibnamefont {Garcia-Santiago}}, \bibinfo
  {author} {\bibfnamefont {C.}~\bibnamefont {Rockstuhl}}, \bibinfo {author}
  {\bibfnamefont {M.}~\bibnamefont {Wegener}},\ and\ \bibinfo {author}
  {\bibfnamefont {I.}~\bibnamefont {Fernandez-Corbaton}},\ }\href
  {https://doi.org/10.1103/PhysRevLett.124.033201} {\bibfield  {journal}
  {\bibinfo  {journal} {Phys. Rev. Lett.}\ }\textbf {\bibinfo {volume} {124}},\
  \bibinfo {pages} {033201} (\bibinfo {year} {2020})}\BibitemShut {NoStop}%
\bibitem [{\citenamefont {Czajkowski}\ and\ \citenamefont
  {Antosiewicz}(2022)}]{czajkowskiLocalBulkCircular2022}%
  \BibitemOpen
  \bibfield  {author} {\bibinfo {author} {\bibfnamefont {K.~M.}\ \bibnamefont
  {Czajkowski}}\ and\ \bibinfo {author} {\bibfnamefont {T.~J.}\ \bibnamefont
  {Antosiewicz}},\ }\href {https://doi.org/10.1515/nanoph-2022-0293} {\bibfield
   {journal} {\bibinfo  {journal} {Nanophotonics}\ }\textbf {\bibinfo {volume}
  {11}},\ \bibinfo {pages} {4287} (\bibinfo {year} {2022})}\BibitemShut
  {NoStop}%
\bibitem [{\citenamefont {Rui}\ \emph {et~al.}(2022)\citenamefont {Rui},
  \citenamefont {Zou}, \citenamefont {Gu},\ and\ \citenamefont
  {Cui}}]{ruiSurfaceEnhancedCircularDichroism2022}%
  \BibitemOpen
  \bibfield  {author} {\bibinfo {author} {\bibfnamefont {G.}~\bibnamefont
  {Rui}}, \bibinfo {author} {\bibfnamefont {S.}~\bibnamefont {Zou}}, \bibinfo
  {author} {\bibfnamefont {B.}~\bibnamefont {Gu}},\ and\ \bibinfo {author}
  {\bibfnamefont {Y.}~\bibnamefont {Cui}},\ }\href
  {https://doi.org/10.1021/acs.jpcc.1c09618} {\bibfield  {journal} {\bibinfo
  {journal} {J. Phys. Chem. C}\ }\textbf {\bibinfo {volume} {126}},\ \bibinfo
  {pages} {2199} (\bibinfo {year} {2022})}\BibitemShut {NoStop}%
\bibitem [{\citenamefont {Liu}\ \emph {et~al.}(2022)\citenamefont {Liu},
  \citenamefont {Deng}, \citenamefont {Guo}, \citenamefont {Yang},
  \citenamefont {Xia}, \citenamefont {Yan}, \citenamefont {Yang}, \citenamefont
  {Qin},\ and\ \citenamefont {Bi}}]{liu_enhanced_2022}%
  \BibitemOpen
  \bibfield  {author} {\bibinfo {author} {\bibfnamefont {W.}~\bibnamefont
  {Liu}}, \bibinfo {author} {\bibfnamefont {L.}~\bibnamefont {Deng}}, \bibinfo
  {author} {\bibfnamefont {Y.}~\bibnamefont {Guo}}, \bibinfo {author}
  {\bibfnamefont {W.}~\bibnamefont {Yang}}, \bibinfo {author} {\bibfnamefont
  {S.}~\bibnamefont {Xia}}, \bibinfo {author} {\bibfnamefont {W.}~\bibnamefont
  {Yan}}, \bibinfo {author} {\bibfnamefont {Y.}~\bibnamefont {Yang}}, \bibinfo
  {author} {\bibfnamefont {J.}~\bibnamefont {Qin}},\ and\ \bibinfo {author}
  {\bibfnamefont {L.}~\bibnamefont {Bi}},\ }\href
  {https://doi.org/10.1364/OE.463918} {\bibfield  {journal} {\bibinfo
  {journal} {Opt. Express, {OE}}\ }\textbf {\bibinfo {volume} {30}},\ \bibinfo
  {pages} {26306} (\bibinfo {year} {2022})}\BibitemShut {NoStop}%
\bibitem [{\citenamefont {Nye}\ and\ \citenamefont
  {Hajnal}(1987)}]{WaveStructureMonochromatic1987}%
  \BibitemOpen
  \bibfield  {author} {\bibinfo {author} {\bibfnamefont {J.~F.}\ \bibnamefont
  {Nye}}\ and\ \bibinfo {author} {\bibfnamefont {J.~V.}\ \bibnamefont
  {Hajnal}},\ }\href {https://doi.org/10.1098/rspa.1987.0002} {\bibfield
  {journal} {\bibinfo  {journal} {Proc. R. Soc. Lond. A}\ }\textbf {\bibinfo
  {volume} {409}},\ \bibinfo {pages} {21} (\bibinfo {year} {1987})}\BibitemShut
  {NoStop}%
\bibitem [{\citenamefont {Dennis}\ \emph {et~al.}(2009)\citenamefont {Dennis},
  \citenamefont {O'Holleran},\ and\ \citenamefont
  {Padgett}}]{dennis_chapter_2009}%
  \BibitemOpen
  \bibfield  {author} {\bibinfo {author} {\bibfnamefont {M.~R.}\ \bibnamefont
  {Dennis}}, \bibinfo {author} {\bibfnamefont {K.}~\bibnamefont {O'Holleran}},\
  and\ \bibinfo {author} {\bibfnamefont {M.~J.}\ \bibnamefont {Padgett}},\ }in\
  \href {https://doi.org/10.1016/S0079-6638(08)00205-9} {\emph {\bibinfo
  {booktitle} {Progress in Optics}}},\ Vol.~\bibinfo {volume} {53}\ (\bibinfo
  {publisher} {Elsevier},\ \bibinfo {year} {2009})\ pp.\ \bibinfo {pages}
  {293--363}\BibitemShut {NoStop}%
\bibitem [{\citenamefont {Peng}\ \emph {et~al.}(2021)\citenamefont {Peng},
  \citenamefont {Liu},\ and\ \citenamefont {Wang}}]{peng_polarization_2021}%
  \BibitemOpen
  \bibfield  {author} {\bibinfo {author} {\bibfnamefont {J.}~\bibnamefont
  {Peng}}, \bibinfo {author} {\bibfnamefont {W.}~\bibnamefont {Liu}},\ and\
  \bibinfo {author} {\bibfnamefont {S.}~\bibnamefont {Wang}},\ }\href
  {https://doi.org/10.1103/PhysRevA.103.023520} {\bibfield  {journal} {\bibinfo
   {journal} {Phys. Rev. A}\ }\textbf {\bibinfo {volume} {103}},\ \bibinfo
  {pages} {023520} (\bibinfo {year} {2021})}\BibitemShut {NoStop}%
\bibitem [{\citenamefont {Jia}\ \emph {et~al.}(2022)\citenamefont {Jia},
  \citenamefont {Peng}, \citenamefont {Cheng},\ and\ \citenamefont
  {Wang}}]{jia_chiral_2022}%
  \BibitemOpen
  \bibfield  {author} {\bibinfo {author} {\bibfnamefont {S.}~\bibnamefont
  {Jia}}, \bibinfo {author} {\bibfnamefont {J.}~\bibnamefont {Peng}}, \bibinfo
  {author} {\bibfnamefont {Y.}~\bibnamefont {Cheng}},\ and\ \bibinfo {author}
  {\bibfnamefont {S.}~\bibnamefont {Wang}},\ }\href
  {https://doi.org/10.1103/PhysRevA.105.033513} {\bibfield  {journal} {\bibinfo
   {journal} {Phys. Rev. A}\ }\textbf {\bibinfo {volume} {105}},\ \bibinfo
  {pages} {033513} (\bibinfo {year} {2022})}\BibitemShut {NoStop}%
\bibitem [{\citenamefont {Peng}\ \emph {et~al.}(2022)\citenamefont {Peng},
  \citenamefont {Zhang}, \citenamefont {Jia}, \citenamefont {Liu},\ and\
  \citenamefont {Wang}}]{peng_topological_2022}%
  \BibitemOpen
  \bibfield  {author} {\bibinfo {author} {\bibfnamefont {J.}~\bibnamefont
  {Peng}}, \bibinfo {author} {\bibfnamefont {R.-Y.}\ \bibnamefont {Zhang}},
  \bibinfo {author} {\bibfnamefont {S.}~\bibnamefont {Jia}}, \bibinfo {author}
  {\bibfnamefont {W.}~\bibnamefont {Liu}},\ and\ \bibinfo {author}
  {\bibfnamefont {S.}~\bibnamefont {Wang}},\ }\href
  {https://doi.org/10.1126/sciadv.abq0910} {\bibfield  {journal} {\bibinfo
  {journal} {Sci. Adv.}\ }\textbf {\bibinfo {volume} {8}},\ \bibinfo {pages}
  {eabq0910} (\bibinfo {year} {2022})}\BibitemShut {NoStop}%
\bibitem [{\citenamefont {Olmon}\ \emph {et~al.}(2012)\citenamefont {Olmon},
  \citenamefont {Slovick}, \citenamefont {Johnson}, \citenamefont {Shelton},
  \citenamefont {Oh}, \citenamefont {Boreman},\ and\ \citenamefont
  {Raschke}}]{olmon_optical_2012}%
  \BibitemOpen
  \bibfield  {author} {\bibinfo {author} {\bibfnamefont {R.~L.}\ \bibnamefont
  {Olmon}}, \bibinfo {author} {\bibfnamefont {B.}~\bibnamefont {Slovick}},
  \bibinfo {author} {\bibfnamefont {T.~W.}\ \bibnamefont {Johnson}}, \bibinfo
  {author} {\bibfnamefont {D.}~\bibnamefont {Shelton}}, \bibinfo {author}
  {\bibfnamefont {S.-H.}\ \bibnamefont {Oh}}, \bibinfo {author} {\bibfnamefont
  {G.~D.}\ \bibnamefont {Boreman}},\ and\ \bibinfo {author} {\bibfnamefont
  {M.~B.}\ \bibnamefont {Raschke}},\ }\href
  {https://doi.org/10.1103/PhysRevB.86.235147} {\bibfield  {journal} {\bibinfo
  {journal} {Phys. Rev. B}\ }\textbf {\bibinfo {volume} {86}},\ \bibinfo
  {pages} {235147} (\bibinfo {year} {2012})}\BibitemShut {NoStop}%
\bibitem [{\citenamefont {Vázquez-Lozano}\ and\ \citenamefont
  {Martínez}(2018)}]{vazquez-lozano_optical_2018}%
  \BibitemOpen
  \bibfield  {author} {\bibinfo {author} {\bibfnamefont {J.~E.}\ \bibnamefont
  {Vázquez-Lozano}}\ and\ \bibinfo {author} {\bibfnamefont {A.}~\bibnamefont
  {Martínez}},\ }\href {https://doi.org/10.1103/PhysRevLett.121.043901}
  {\bibfield  {journal} {\bibinfo  {journal} {Phys. Rev. Lett.}\ }\textbf
  {\bibinfo {volume} {121}},\ \bibinfo {pages} {043901} (\bibinfo {year}
  {2018})}\BibitemShut {NoStop}%
\end{thebibliography}%
\end{document}